# New Designs on MVDR Robust Adaptive Beamforming Based on Optimal Steering Vector Estimation

Yongwei Huang, Mingkang Zhou, Sergiy A. Vorobyov

*Abstract*—The robust adaptive beamforming design problem based on estimation of the signal of interest steering vector is considered in the paper. In this case, the optimal beamformer is obtained by computing the sample matrix inverse and an optimal estimate of the signal of interest steering vector. The common criteria to find the best estimate of the steering vector are the beamformer output signal-to-noise-plus-interference ratio (SINR) and output power, while the constraints assume as little as possible prior inaccurate knowledge about the signal of interest, the propagation media, and the antenna array. Herein, in order to find the optimal steering vector estimate of the signal of interest, a new beamformer output power maximization problem is formulated and solved subject to a double-sided norm perturbation constraint, a similarity constraint, and a quadratic constraint that guarantees that the direction-of-arrival (DOA) of the signal of interest is away from the DOA region of all linear combinations of the interference steering vectors. In the new robust design, the prior information required consists of some allowable error norm bounds, the approximate knowledge of the antenna array geometry, and the angular sector of the signal of interest. It turns out that the array output power maximization problem is a non-convex quadratically constrained quadratic programming problem with inhomogeneous constraints. However, we show that the problem is still solvable, and develop efficient algorithms for finding globally optimal estimate of the signal of interest steering vector. The results are generalized to the case where an ellipsoidal constraint (rather than the similarity constraint) is considered, and sufficient conditions for the global optimality are derived. In addition, a new quadratic constraint on the actual signal steering vector is proposed in order to improve the array performance. To validate our results, simulation examples are presented, and they demonstrate the improved performance of the new robust beamformers in terms of the output SINR as well as the output power.

## I. Introduction

For decades array signal processing has been wildly employed in many applications to radar, sonar, communications, microphone array speech/audio processing [1] to mention just a few best known. In array processing, robust adaptive beamforming has particularly been recognized as a fundamental problem and drawn much research interest. It is because the traditional techniques of adaptive beamforming such as Capon beamforming method have weak immunity against small or modest differences between the presumed and actual signal steering vectors [2], pointing and antenna calibration errors, and others mismatches [3], [4]. Therefore, to meet the robustness demands, a number of robust adaptive beamforming techniques have been established, and substantial progress in the areas has been made especially in the last two decades, partially supported by significant developments in convex and robust optimization [5].

Let us first introduce the basic concepts and the most recent developments in robust adaptive beamforming that this work will continue on. We start with introducing the notations that will be used throughout the paper. We adopt the notation of using boldface for vectors $\boldsymbol{a}$ (lower case), and matrices $\boldsymbol{A}$ (upper case). The transpose operator and the conjugate transpose operator are denoted by the symbols $(\cdot)^T$ and $(\cdot)^H$ respectively. The notation $\mathrm{tr}(\cdot)$ stands for the trace of the square matrix argument; $\boldsymbol{I}$ and $\boldsymbol{0}$ denote respectively the identity matrix and the matrix (or the row vector or the column vector) with zero entries (their size is determined from the context). The letter $j$ represents the imaginary unit (i.e. $j = \sqrt{-1}$), while the letter $i$ often serves as index in this paper. For any complex number $x$, we use $\Re(x)$ and $\Im(x)$ to denote respectively the real and the imaginary parts of $x$, $|x|$ and $\arg(x)$ represent the modulus and the argument of $x$, and $x^*$ ($\boldsymbol{x}^*$ or $\boldsymbol{X}^*$) stands for the (component-wise) conjugate of $x$ ($\boldsymbol{x}$ or $\boldsymbol{X}$). The Euclidean norm (the Frobenius norm) of the vector $\boldsymbol{x}$ (the matrix $\boldsymbol{X}$) is denoted by $\|\boldsymbol{x}\|$ ($\|\boldsymbol{X}\|$). The curled inequality symbol $\succeq$ (and its strict form $\succ$) is used to denote generalized inequality: $\boldsymbol{A} \succeq \boldsymbol{B}$ means that $\boldsymbol{A} - \boldsymbol{B}$ is an Hermitian positive semidefinite matrix ($\boldsymbol{A} \succ \boldsymbol{B}$ for positive definiteness). The space of Hermitian $N \times N$ matrices (the space of real-valued symmetric $N \times N$ matrices) is denoted by $\mathcal{H}^N$ ($\mathcal{S}^N$), and the set of all positive semidefinite matrices in $\mathcal{H}^N$ ($\mathcal{S}^N$) by $\mathcal{H}^N_+$ ($\mathcal{S}^N_+$). $\mathsf{E}[\cdot]$ represents the statistical expectation. Finally, $v^\star(\cdot)$ represents the optimal value of an optimization problem.

Let us consider a receive narrowband beamformer applied to an output of a linear array of $N$ antenna elements. The output signal of the beamformer at the time instant $k$ can be written as

$$y(k) = \boldsymbol{w}^H \boldsymbol{x}(k), \qquad (1)$$

where $\boldsymbol{w}$ is the $N \times 1$ vector of complex weight coefficients, i.e., the beamvector, and $\boldsymbol{x}(k)$ is the complex vector of the antenna array measurements. The array observation vector in (1) is given by

$$\boldsymbol{x}(k) = \boldsymbol{s}(k) + \boldsymbol{i}(k) + \boldsymbol{n}(k), \qquad (2)$$

where $\boldsymbol{s}(k)$, $\boldsymbol{i}(k)$, and $\boldsymbol{n}(k)$ are statistically independent vectors corresponding, respectively, to the signal of interest, interference and sensor noise. The signal of interest can be written under the point source assumption as $\boldsymbol{s}(k) = s(k)\boldsymbol{a}$, where $s(k)$ is the signal waveform and $\boldsymbol{a}$ is the steering vector.

The optimal weight vector $\boldsymbol{w}^\star$ can be found from the optimal solution of the following signal-to-interference-plus-noise ratio (SINR) maximization problem

$$\underset{\boldsymbol{w}}{\text{maximize}}\, \text{SINR} = \underset{\boldsymbol{w}}{\text{maximize}}\, \frac{\sigma_s^2 |\boldsymbol{w}^H \boldsymbol{a}|^2}{\boldsymbol{w}^H \boldsymbol{R}_{i+n} \boldsymbol{w}} \qquad (3)$$

where $\sigma_s^2$ is the signal of interest power and $\boldsymbol{R}_{i+n} = \mathsf{E}[(\boldsymbol{i}(k) + \boldsymbol{n}(k))(\boldsymbol{i}(k) + \boldsymbol{n}(k))^H]$ is the interference-plus-noise covariance matrix. Since the exact covariance matrix $\boldsymbol{R}_{i+n}$ is unknown in practice, the following sample data covariance matrix computed based on $T$ available snapshots

$$\hat{\boldsymbol{R}} = \frac{1}{T} \sum_{k=1}^{T} \boldsymbol{x}(k) \boldsymbol{x}^H(k) \qquad (4)$$

often is employed instead of $\boldsymbol{R}_{i+n}$ in the SINR maximization problem (3). It is evident that the SINR maximization problem is tantamount to the following optimization problem

$$\underset{\boldsymbol{w}}{\text{minimize}}\, \boldsymbol{w}^H \hat{\boldsymbol{R}} \boldsymbol{w} \quad \text{subject to} \quad |\boldsymbol{w}^H \boldsymbol{a}| = 1 \qquad (5)$$

Y. Huang and M. Zhou are with School of Information Engineering, Guangdong University of Technology, University Town, Guangzhou, Guangdong, China. E-mail: ywhuang@gdut.edu.cn.
S.A. Vorobyov is with Department of Signal Processing and Acoustics, Aalto University, Konemiehentie 2, 02150 Espoo, Finland. E-mail: svor@ieee.org.

with the optimal solution

$$\boldsymbol{w}^\star = \frac{1}{\boldsymbol{a}^H \hat{\boldsymbol{R}}^{-1} \boldsymbol{a}} \hat{\boldsymbol{R}}^{-1} \boldsymbol{a}, \qquad (6)$$

referred to as minimum variance distortionless response (MVDR) sampling matrix invert (SMI) beamformer [1] or Capon beamformer [2]. Let us also note here that the corresponding array output power $\mathsf{E}[|\boldsymbol{y}(k)|^2]$ is given by

$$\mathsf{E}[|\boldsymbol{y}(k)|^2] = \mathsf{E}[|\boldsymbol{w}^{\star H} \boldsymbol{x}(k)|^2] \approx \boldsymbol{w}^{\star H} \hat{\boldsymbol{R}} \boldsymbol{w}^\star = \frac{1}{\boldsymbol{a}^H \hat{\boldsymbol{R}}^{-1} \boldsymbol{a}}. \qquad (7)$$

In practice, the desired signal steering vector $\boldsymbol{a}$ is usually known imprecisely, while only some presumed steering vector $\hat{\boldsymbol{a}}$ can be estimated based on the knowledge of antenna array geometry, parameters of the signal of interest, and also some additional assumptions about propagation media and antenna array calibration. As a result, in many practical scenarios, the performance of the beamformer (6) degrades dramatically because of the mismatch between the actual steering vector $\boldsymbol{a}$ and the presumed steering vector $\hat{\boldsymbol{a}}$, as well as the inaccurate estimate $\hat{\boldsymbol{R}}$. To mitigate the degradation, a number of robust adaptive beamforming techniques based on the modeling of the steering vector mismatch as an additive deterministic norm bounded vector have been proposed in the last two decades (see [2], [3], [4], [6], [7], [8], [9] and references therein). If the presumed steering vector and especially the bound on the mismatch norm are hard to estimate in practice, the alternative robust adaptive beamforming design has been obtained based on stochastic model for the steering vector mismatch by requesting the distortionless response constraint to be satisfied with high (practically acceptable) probability [10]. With respect of the estimate $\hat{\boldsymbol{R}}$, the worst-case approach results in the diagonal loading of the data covariance matrix sample estimate. Besides, many approaches to robust estimation of the data covariance matrix have been developed as well. The most notable are the random matrix theory-based techniques, subspace techniques, Bayesian techniques, shrinkage techniques, and covariance matrix reconstruction techniques [11], [12], [13], [14], [15], [16], [17]. However, the typical theme in these developments is to assume more prior information for obtaining better robust adaptive beamforming designs. It goes partially against the motivation for robust designs, which is to guarantee a reliable performance (acceptably high output SINR) with as little as possible prior information as it is extensively argued and studied in [18], [19], [20]. The only prior information used in [18] is the imprecise knowledge of the angular sector of the signal of interest and antenna array geometry, while the knowledge of the presumed steering vector is not needed as was first studied in [21].

Summarizing, in the aforementioned approach, the MVDR robust adaptive beamformer adopts the beamvector (6) with $\boldsymbol{a}$ therein replaced by an estimate $\hat{\boldsymbol{a}}$ that is optimized via a certain method, while assuming that $\hat{\boldsymbol{R}}$ is a sufficiently good estimate of $\boldsymbol{R}$. In [18], the optimal steering vector $\hat{\boldsymbol{a}}$ is picked up by maximizing the beamformer output power (7) subject to a constraint separating the direction of arrival (DOA) of the signal of interest from the directions given by linear combinations of the interference steering vectors, as well as a norm constraint of the steering vector. Mathematically, using (7) as an objective, the array output power maximization problem of finding an optimal $\hat{\boldsymbol{a}}$ is cast as

$$\begin{array}{ll} \underset{\boldsymbol{a}}{\text{minimize}} & \boldsymbol{a}^H \hat{\boldsymbol{R}}^{-1} \boldsymbol{a} \\ \text{subject to} & \boldsymbol{a}^H \tilde{\boldsymbol{C}} \boldsymbol{a} \leq \Delta_0 \\ & \|\boldsymbol{a}\|^2 = N, \end{array} \qquad (8)$$

where

$$\tilde{\boldsymbol{C}} = \int_{\tilde{\Theta}} \boldsymbol{d}(\theta) \boldsymbol{d}^H(\theta) d\theta, \qquad (9)$$

and $\boldsymbol{d}(\theta)$ is the steering vector associated with $\theta$ that has the structure defined by the antenna array geometry. In (9), $\tilde{\Theta}$ is the complement of the angular sector $\Theta = [\theta_{\min}, \theta_{\max}]$, in which the direction of the signal of interest lies, and it is assumed to be separated from general locations of the interfering signals. Also, the parameter $\Delta_0$ is obtained by

$$\Delta_0 = \max_{\theta \in \Theta} \boldsymbol{d}^H(\theta) \tilde{\boldsymbol{C}} \boldsymbol{d}(\theta). \qquad (10)$$

In fact, $\Delta_0$ is a boundary line to distinguish approximately whether or not the direction of $\boldsymbol{a}$ is in the actual signal angular sector $\Theta$. Specifically, if

$$\boldsymbol{a}^H \tilde{\boldsymbol{C}} \boldsymbol{a} \leq \Delta_0, \qquad (11)$$

the direction of $\boldsymbol{a}$ is treated as being inside $\Theta$, which means that the direction of $\boldsymbol{a}$ never converges to the direction of any steering vector associated with a linear combination of the interferers (cf. [18, (23)-(25)]), and see also Fig. 2 in [18] or Fig. 1(b) below for reference. Note that the definition of the matrix $\tilde{\boldsymbol{C}}$ requires only the knowledge about the antenna array geometry (relating to $\boldsymbol{d}(\theta)$) and the angular sector $\Theta$ and $\Delta_0$ can be easily found based on (10). No other prior knowledge is required.

As for how to solve (8), observe that problem (8) is a homogeneous quadratically-constrained quadratic programming (QCQP) problem with two constraints only, and thus its optimal solution can be found efficiently through solving its semi-definite programming (SDP) relaxed problem followed by a procedure of retrieving a rank-one solution from a general rank relaxed solution when necessary (see e.g. [22], [23], [24], [25], [26]. Specifically for problem (8), thanks to the structured matrices there, the procedure of finding a rank-one solution can be simplified significantly as stated in [18, Theorem 1]. The computational burden is dominated by solving the SDP relaxation problem, the worst-case complexity of which is $O(N^{4.5})$ (cf. [26], [27]).

In this paper, we aim to provide an optimal design with improved performance for MVDR robust adaptive beamforming based on signal steering vector estimation to that of the state-of-the-art design (8). The performance is evaluated based on the beamformer output SINR and the beamformer output power. The higher they are, the better. On the other hand, the requirement of a prior information for the robust design are the less, the better, since the prior knowledge often is imprecise and incomplete, which affects the performance of the robust beamformer. In particular, we propose here a new robust adaptive beamforming design with improved performance by introducing more practical constraints to power maximization problem (8), beyond the quadratic constraint setting the DOA of the signal of interest steering vector apart from the DOA interval of all interference sources and their combinations.

First, we extend the steering vector norm equality constraint to a double-sided constraint, allowing a certain range of the error norm perturbations. The new constraint is to account for the steering vector gain perturbations caused, e.g., by the sensor amplitude errors, phase errors as well as the sensor position errors (cf. [6, pp. 2408 and 2414]).

Second, we add a similarity constraint, making sure that the optimal estimate, in terms of the norm of difference, is sufficiently close to at least one steering vector with its DOA inside the angular sector of the signal of interest. The relaxed norm constraint together with the additional similarity constraint in the new problem permits possible enlargement of the feasible set of the problem, and thus, allows to search for a better estimate of the true steering vector. In the new robust design, the only possibly additional prior information includes three allowable error norm bounds.

The formulated optimization problem for the MVDR robust adaptive beamforming is a non-convex QCQP problem with three inho-mogeneous constraints. In general, it may not be solvable, and the scenario is similar to that of some hard QCQP subproblems in a trust-region algorithm in nonlinear programming. However, by some manipulations including a proof of the equivalence between a linear constraint and a quadratic constraint, we show that this non-convex QCQP problem is still solvable and equivalent to its SDP relaxed problem.

Third, we generalize the above results to the case where an ellipsoidal constraint on the signal of interest steering vector, instead of the similarity constraint, is considered. However, the formulated new QCQP problem is a hard problem, and there is no global optimality guaranteed. As a compromise, we thus establish sufficient conditions (easily verifiable) for the global optimality. Although one observes that more prior information about parameters of the ellipsoid is required for the ellipsoidal constraint, better performance of the corresponding robust beamformer can be expected, as a trade-off.

Fourth, a new quadratic constraint is proposed in the optimal estimation problem of the actual signal steering vector. In addition to (11), the constraint provides a new benchmark line setting the direction of the signal of interest steering vector apart from the direction set of linear combinations of all interference steering vectors. With the constraint in hand, the corresponding optimal steering vector estimation problems are built, aiming to provide a new optimal estimate of the signal of interest steering vector, which leads to improvement of the array output performance.

This paper is organized as follows. In Section II, we give several new formulations for robust adaptive beamforming based on maximization of the output power, and describe how to efficiently solve the problems in Section III. A new quadratic constraint for the signal of interest steering vector and the corresponding optimization problems of robust adaptive beamforming design are studied in Section IV. Numerical examples showing the improved performance of the new robust adaptive beamformers are given in Section V. Our conclusions are made in Section VI. Some proofs which are not directly used for developments in the paper are given in Appendix.

## II. NEW IMPROVED ROBUST ADAPTIVE BEAMFORMING DESIGNS

In this section, we formulate a new robust adaptive beamforming problem with the objective to improve the performance of the robust adaptive beamformers from the class considered in this paper, which has been previously shown to be best performing.

In order to ameliorate the array output SINR and output power of the robust adaptive beamformer of (8), a practical way of compromise between sensitivity and robustness is to acquire a bit of more prior information than that required by (8), and formulate new optimal estimation problems of the signal of interest steering vector. There are a number of works addressing such compromise, e.g., doubly constrained beamformer [6], [28], quadratically constrained beamformer [8], beamformer with a general convex uncertainty set of the signal of interest directions [9], and some others mentioned in the survey paper [3].

Herein, we consider the following general steering vector estimation problem:

$$\begin{aligned} \underset{\boldsymbol{a}}{\text{minimize}} \quad & \boldsymbol{a}^H \hat{\boldsymbol{R}}^{-1} \boldsymbol{a} \\ \text{subject to} \quad & \boldsymbol{a}^H \tilde{\boldsymbol{C}} \boldsymbol{a} \leq \Delta_0 \\ & \boldsymbol{a} \in \mathcal{A}, \end{aligned} \qquad (12)$$

where $\mathcal{A}$ stands for the uncertainty set of steering vector $\boldsymbol{a}$. The most common $\mathcal{A}$ includes norm constraint $\|\boldsymbol{a}\|^2 = N$ and others whenever necessary. In particular, when

$$\mathcal{A}_0 = \{\boldsymbol{a} \mid \|\boldsymbol{a}\|^2 = N\} \qquad (13)$$

problem (12) corresponds to problem (8).

In this paper, we consider the uncertainty set defined by the following two conditions:

$$\begin{aligned} \mathcal{A} = \{\boldsymbol{a} \mid & N(1-\eta_1) \leq \|\boldsymbol{a}\|^2 \leq N(1+\eta_2) \\ & \|\boldsymbol{Q}^H(\boldsymbol{a}-\boldsymbol{a}_0)\|^2 \leq \epsilon\}. \end{aligned} \qquad (14)$$

where $\boldsymbol{a}_0 = \boldsymbol{d}(\theta_0)$ and $\theta_0 = (\theta_{\max} + \theta_{\min})/2$ is the middle value of the region $\Theta$. Our purpose is to find a globally optimal solution for (12) with $\mathcal{A}$ in (14). In that uncertainty set, the first double-sided norm constraint accounts for gain perturbations of the array response vector $\boldsymbol{a}$ (see e.g. [2, Sec. 3.2]). The generalized similarity condition in (14) implies that imperfect knowledge of the desired vector $\boldsymbol{a}$ is described as in a convex set (in particular, an ellipsoidal set when $\boldsymbol{Q}$ is of full row rank, in other words, $\boldsymbol{Q}\boldsymbol{Q}^H$ is invertible).

It follows that problem (12) can be formulated as the following QCQP:

$$\begin{aligned} \underset{\boldsymbol{a}}{\text{minimize}} \quad & \boldsymbol{a}^H \hat{\boldsymbol{R}}^{-1} \boldsymbol{a} \\ \text{subject to} \quad & \boldsymbol{a}^H \tilde{\boldsymbol{C}} \boldsymbol{a} \leq \Delta_0 \\ & N(1-\eta_1) \leq \|\boldsymbol{a}\|^2 \leq N(1+\eta_2) \\ & \|\boldsymbol{Q}^H(\boldsymbol{a}-\boldsymbol{a}_0)\|^2 \leq \epsilon. \end{aligned} \qquad (15)$$

Note that we assume throughout that $\boldsymbol{Q}^H \boldsymbol{a}_0 \neq \boldsymbol{0}$; otherwise, $\boldsymbol{Q}^H \boldsymbol{a}_0$ vanishes in the generalized similarity constraint and the constraint becomes $\|\boldsymbol{Q}^H \boldsymbol{a}\|^2 \leq \epsilon$, which makes the similarity constraint meaningless. QCQP problem (15) has one double-sided constraint, one homogeneous and one more inhomogeneous inequality constraints. Although it follows from [22], [23] that only conditional optimality[1] can be achieved for the problems of type (15), our purpose herein is to identify such instances for the double-sided QCQP problem (15) when it can be solved up to the global optimality.

Particularly, when $\boldsymbol{Q}$ is the identity matrix, namely, $\boldsymbol{Q} = \boldsymbol{I}$, the third condition in (15) becomes a sphere constraint, or just a traditional similarity constraint. Problem (15) can be then rewritten into:

$$\begin{aligned} \underset{\boldsymbol{a}}{\text{minimize}} \quad & \boldsymbol{a}^H \hat{\boldsymbol{R}}^{-1} \boldsymbol{a} \\ \text{subject to} \quad & \boldsymbol{a}^H \tilde{\boldsymbol{C}} \boldsymbol{a} \leq \Delta_0 \\ & N(1-\eta_1) \leq \|\boldsymbol{a}\|^2 \leq N(1+\eta_2) \\ & \|\boldsymbol{a}-\boldsymbol{a}_0\|^2 \leq \epsilon. \end{aligned} \qquad (16)$$

Comparing problem (16) with problem (8), it can be seen that the additional prior knowledge required in (16) includes the parameter $\boldsymbol{a}_0$. However, $\boldsymbol{a}_0$ is set by default to $\boldsymbol{d}(\theta_0)$ with $\theta_0$ being the middle point of the desired sector $\Theta$. Therefore, it is sufficient to have the knowledge of $\Theta$ and $\boldsymbol{d}(\theta)$ in order to know $\boldsymbol{a}_0$. We thus highlight that the prior knowledge required in (16) includes those in (8) (namely the angular sector $\Theta$ of interest and the approximate knowledge of antenna array geometry), and $\eta_1$, $\eta_2$ and $\epsilon$, which are user parameters and are just allowable norm error bounds.

## III. SOLVING PROBLEMS (15) AND (16)

### A. Solving problem (16), $\boldsymbol{Q} = \boldsymbol{I}$ in (15)

In this subsection, we show that the global optimality for (15) can be achieved when $\boldsymbol{Q} = \boldsymbol{I}$, and consequently, a solution procedure of polynomial time is devised. In (16), the ellipsoid constraint of (15) reduces to a ball (centered at $\boldsymbol{a}_0$) constraint, which simplifies the problem and allows for the global optimality as it is summarised in the follow theorem. We give the theorem with its proof here because the solution method follows from the proof.

---

[1]It means that the global optimal value can be attained only under some sufficient conditions.

We first cite a rank-one matrix decomposition lemma, which will be employed soon.

**Lemma III.1 (Theorem 2.1 in [23])** *Suppose that $X$ is an $N \times N$ complex Hermitian positive semidefinite matrix of rank $R$, and $A$, $B$ are two given $N \times N$ Hermitian matrices. Then, there exists a rank-one decomposition $X = \sum_{r=1}^{R} x_r x_r^H$ such that*

$$x_r^H A x_r = \frac{\operatorname{tr}(AX)}{R} \quad \text{and} \quad x_r^H B x_r = \frac{\operatorname{tr}(BX)}{R}, \ r = 1, \ldots, R.$$

The rank-one decomposition synthetically is denoted as $\{x_r\} = \mathcal{D}_1(X, A, B)$.

The global optimality for (16) is given as follows.

**Theorem III.2** *QCQP problem* (16) *is solvable and its globally optimal solution can be returned tractably within polynomial time.*

*Proof:* Let us consider the following SDP relaxation problem:

$$\begin{aligned}
\underset{X, y}{\text{minimize}} \quad & \operatorname{tr}(\hat{R}^{-1} X) \\
\text{subject to} \quad & \operatorname{tr}(\tilde{C} X) \leq \Delta_0 \\
& N(1 - \eta_1) \leq \operatorname{tr} X \leq N(1 + \eta_2) \\
& \operatorname{tr}\left(\begin{bmatrix} I & -a_0 \\ -a_0^H & N \end{bmatrix} \begin{bmatrix} X & y \\ y^H & 1 \end{bmatrix}\right) \leq \epsilon, \\
& \begin{bmatrix} X & y \\ y^H & 1 \end{bmatrix} \succeq 0
\end{aligned} \quad (17)$$

Clearly it is solvable since the objective is continuous and the feasible set is compact. Suppose that $(X^\star, y^\star)$ is an optimal solution which can be computed via an interior-point method (see e.g. [27]), and let $b_1 = \operatorname{tr}(\tilde{C} X^\star)$ and $b_2 = \operatorname{tr}(X^\star)$. Therefore, (17) is equivalent to

$$\begin{aligned}
\underset{X, y}{\text{minimize}} \quad & \operatorname{tr}(\hat{R}^{-1} X) \\
\text{subject to} \quad & \operatorname{tr}(\tilde{C} X) = b_1, \\
& \operatorname{tr}(X) = b_2, \\
& \operatorname{tr}\left(\begin{bmatrix} I & -a_0 \\ -a_0^H & N \end{bmatrix} \begin{bmatrix} X & y \\ y^H & 1 \end{bmatrix}\right) \leq \epsilon, \\
& \begin{bmatrix} X & y \\ y^H & 1 \end{bmatrix} \succeq 0,
\end{aligned} \quad (18)$$

which is a conventional SDP relaxation problem for the QCQP problem:

$$\begin{aligned}
\underset{a}{\text{minimize}} \quad & a^H \hat{R}^{-1} a \\
\text{subject to} \quad & a^H \tilde{C} a = b_1, \\
& \|a\|^2 = b_2, \\
& \|a - a_0\|^2 \leq \epsilon.
\end{aligned} \quad (19)$$

Note that the third constraint can be simplified into

$$\Re(a_0^H a) \geq (b_2 + N - \epsilon)/2, \quad (20)$$

which always amounts to the constraint

$$|a_0^H a| \geq (b_2 + N - \epsilon)/2. \quad (21)$$

Indeed, suppose that $a$ complies with (20), it is evident that $a$ satisfies (21) too. Conversely, if $a$ satisfies (21), then the phase rotation of $a$:

$$a e^{-j \arg(a_0^H a)} \quad (22)$$

fulfills

$$\Re(a_0^H a e^{-j \arg(a_0^H a)}) = |a_0^H a| \geq (b_2 + N - \epsilon)/2, \quad (23)$$

namely, $a e^{-j \arg(a_0^H a)}$ conforms to (20), but does not alter all other constraints and the objective function values.

Accordingly, one claims that QCQP problem (19) is equivalent to the following QCQP problem:

$$\begin{aligned}
\underset{a}{\text{minimize}} \quad & a^H \hat{R}^{-1} a \\
\text{subject to} \quad & a^H \tilde{C} a = b_1, \\
& \|a\|^2 = b_2, \\
& a^H a_0 a_0^H a \geq ((b_2 + N - \epsilon)/2)^2.
\end{aligned} \quad (24)$$

Again the SDP relaxation problem of the QCQP is expressed as:

$$\begin{aligned}
\underset{X}{\text{minimize}} \quad & \operatorname{tr}(\hat{R}^{-1} X) \\
\text{subject to} \quad & \operatorname{tr}(\tilde{C} X) = b_1, \\
& \operatorname{tr} X = b_2, \\
& \operatorname{tr}(a_0 a_0^H X) \geq ((b_2 + N - \epsilon)/2)^2, \\
& X \succeq 0.
\end{aligned} \quad (25)$$

It follows from [22], [23], [29] that SDP problem (25) has always a rank-one solution (since it is solvable), and that the rank-one solution can be returned by the matrix decomposition described in Lemma III.1. Therefore, we have $v^\star((24)) = v^\star((25))$. In fact, suppose that $X^\star$ is optimal for (25), and let $b_3 = \operatorname{tr}(a_0 a_0^H X^\star)$. We conduct the rank-one matrix decomposition, getting a vector:

$$x = \mathcal{D}_1\left(X^\star, \tilde{C} - \frac{b_1}{b_2} I, a_0 a_0^H - \frac{b_3}{b_2} I\right) \quad (26)$$

and set

$$x^\star := \frac{\sqrt{b_2} x}{\|x\|}. \quad (27)$$

It is verified that $x^\star x^{\star H}$ is feasible and optimal for (25) (since $x^\star \in \operatorname{Range}(X^\star)$, cf. [29, Theorem 6.6]).

Observe that

$$v^\star((17)) = v^\star((18)) \geq v^\star((19)) = v^\star((24)) = v^\star((25)). \quad (28)$$

Note further that for any optimal solution $(X^\star, y^\star)$ for (17) (as well as (18)), the component $X^\star$ is feasible for (25), since

$$\begin{aligned}
\operatorname{tr}(X^\star a_0 a_0^H) & \geq \operatorname{tr}(y^\star y^{\star H} a_0 a_0^H) \\
& = |y^{\star H} a_0|^2 \\
& \geq (\Re(y^{\star H} a_0))^2 \\
& \geq ((b_2 + N - \epsilon)/2)^2,
\end{aligned}$$

where the first inequality is due to $X^\star \succeq y^\star y^{\star H}$, and the last inequality is due to the third constraints in (18). Therefore, $X^\star$ is optimal for (25) since $\operatorname{tr}(\hat{R}^{-1} X^\star) = v^\star((25))$ (On one hand, $v^\star((17)) = \operatorname{tr}(\hat{R}^{-1} X^\star) \geq v^\star((25))$; on the other hand, $v^\star((25)) \geq \operatorname{tr}(\hat{R}^{-1} X^\star)$ because $X^\star$ is feasible for (25)). Thereby, we have $v^\star((17)) = v^\star((25))$, and it follows from (28) that all optimal values in (28) are equal to each other :

$$v^\star((17)) = v^\star((18)) = v^\star((19)) = v^\star((24)) = v^\star((25)). \quad (29)$$

It follows from the equality chain (29) that the facts below are implied in order.

1) First, suppose $(X^\star, y^\star)$ is optimal for (17), then $X^\star$ is optimal for (25). Perform the matrix decomposition as in (26) and obtain $x^\star$ as in (27). Then $x^\star x^{\star H}$ is optimal for (25) and $x^\star$ is also the optimal solution for (24).
2) Second, conduct the phase rotation $a^\star = x^\star e^{-j \arg(a_0^H x^\star)}$. It follows that $a^\star$ is optimal for (19) and

$$\begin{bmatrix} a^\star a^{\star H} & a^\star \\ a^{\star H} & 1 \end{bmatrix}$$

is a rank-one optimal solution for (18), as well as for (17).
3) Finally, $a^\star$ is optimal for (16).

It completes the proof. ∎

Based on the constructive procedure in the proof of Theorem III.2 (the three facts summarized), we devise the algorithm for finding the optimal solution of (16) as in Algorithm 1.

---
**Algorithm 1** Procedure for Solving QCQP Problem (16)
---
**Input:** $\hat{R}, \tilde{C}, a_0, \Delta_0, \eta, \epsilon$;
**Output:** An optimal solution $a^\star$ of problem (16);
1: solve its SDP relaxation problem (17) and find $(X^\star, y^\star)$;
2: perform the rank-one matrix decomposition and obtain $x = \mathcal{D}_1\left(X^\star, \tilde{C} - \frac{b_1}{b_2}I, a_0 a_0^H - \frac{b_3}{b_2}I\right)$, where $b_1 = \text{tr}(\tilde{C}X^\star)$, $b_2 = \text{tr}(X^\star)$, and $b_3 = \text{tr}(a_0 a_0^H X^\star)$;
3: set $x^\star = \sqrt{b_2}\frac{x}{\|x\|}$;
4: output $a^\star = x^\star e^{-j \arg(a_0^H x^\star)}$.

---

Since the rank-one matrix decomposition $\mathcal{D}_1$ has computational complexity of $O(N^3)$, which is smaller than that of solving the SDP problem, hence the computational cost for Algorithm 1 is dominated by solving the SDP problem (cf. [26]).

### B. Solving problem (15), general $Q$

In the case of general $Q$, the additional prior information including in the matrix $Q$ (it is an ellipsoidal parameter if $Q$ is of full row rank) is required. The reason for considering this more general case is to show the possibility of performance improvement due to the additional prior information in $Q$ by numerical simulations in Section V.

In this subsection, we aim to develop a polynomial algorithm for solving (15) at least suboptimally, but first, we find sufficient conditions under which non-convex problem (15) can be solved up to the global optimality. Toward this end, let us recast (15) into the following equivalent homogeneous QCQP problem:

$$\begin{aligned}
\underset{y}{\text{minimize}} \quad & \text{tr}(A_0 yy^H) \\
\text{subject to} \quad & \text{tr}(A_1 yy^H) \leq \Delta_0 \\
& N(1-\eta_1) \leq \text{tr}(A_2 yy^H) \leq N(1+\eta_2) \\
& \text{tr}(A_3 yy^H) \leq \epsilon \\
& \text{tr}(A_4 yy^H) = 1,
\end{aligned} \quad (30)$$

where $y$ is the vector $a$ augmented with $t$, that is, $y = [a; t] \in \mathbb{C}^{N+1}$,

$$A_0 = \begin{bmatrix} \hat{R}^{-1} & 0 \\ 0 & 0 \end{bmatrix}, A_1 = \begin{bmatrix} \tilde{C} & 0 \\ 0 & 0 \end{bmatrix}, A_2 = \begin{bmatrix} I & 0 \\ 0 & 0 \end{bmatrix}, \quad (31)$$

$$\begin{aligned}
A_3 &= \begin{bmatrix} Q \\ -a_0^H Q \end{bmatrix} [Q^H, -Q^H a_0] \\
&= \begin{bmatrix} QQ^H & -QQ^H a_0 \\ -a_0^H QQ^H & a_0^H QQ^H a_0 \end{bmatrix},
\end{aligned} \quad (32)$$

and

$$A_4 = \begin{bmatrix} 0 & 0 \\ 0 & 1 \end{bmatrix}. \quad (33)$$

It can be easily verified that problems (15) and (30) have the same optimal value:

$$v^\star((15)) = v^\star((30)), \quad (34)$$

and if $y^\star = [x^\star; t^\star]$ is the optimal solution of (30), then $x^\star/t^\star$ is the optimal solution of (15). Therefore, we only need to focus on problem (30) in order to solve (15). Accordingly, we study the SDP relaxation for (30):

$$\begin{aligned}
\underset{Y}{\text{minimize}} \quad & \text{tr}(A_0 Y) \\
\text{subject to} \quad & \text{tr}(A_1 Y) \leq \Delta_0 \\
& N(1-\eta_1) \leq \text{tr}(A_2 Y) \leq N(1+\eta_2) \\
& \text{tr}(A_3 Y) \leq \epsilon \\
& \text{tr}(A_4 Y) = 1 \\
& Y \succeq 0.
\end{aligned} \quad (35)$$

The dual problem of (35) is the following SDP problem:

$$\begin{aligned}
\underset{\{z_i\}}{\text{maximize}} \quad & \Delta_0 z_1 + N(1-\eta_1)z_0 + N(1+\eta_2)z_2 + \epsilon z_3 + z_4 \\
\text{subject to} \quad & A_0 - z_1 A_1 - (z_0 + z_2)A_2 - z_3 A_3 - z_4 A_4 \succeq 0 \\
& z_0 \geq 0, z_1 \leq 0, z_2 \leq 0, z_3 \leq 0, z_4 \in \mathbb{R}.
\end{aligned} \quad (36)$$

It is known (cf. [27]) that the optimality conditions for the primal and dual SDP problems (termed also complementary conditions) are:

$$0 = \text{tr}((A_0 - z_1 A_1 - (z_0 + z_2)A_2 - z_3 A_3 - z_4 A_4)Y) \quad (37)$$
$$0 = z_1(\text{tr}(A_1 Y) - \Delta_0) \quad (38)$$
$$0 = z_0(\text{tr}(A_2 Y) - N(1-\eta_1)) \quad (39)$$
$$0 = z_2(\text{tr}(A_2 Y) - N(1+\eta_2)) \quad (40)$$
$$0 = z_3(\text{tr}(A_3 Y) - \epsilon) \quad (41)$$

Let us establish now several sufficient conditions under which SDP problem (35) possesses a rank-one solution. With this objective in mind, suppose that both primal SDP problem (35) and dual SDP problem (36) are solvable, and let $(Y^\star; \{z_i^\star\})$ denote the optimal primal-dual pair. The solution $Y^\star$ is assumed to have the form:

$$Y^\star = \begin{bmatrix} X^\star & x^\star \\ x^{\star H} & 1 \end{bmatrix}. \quad (42)$$

Then the following theorem tells when the solution of the form (42) is rank-one solution of SDP problem (35). The proof of the theorem is also given because the constructions introduced in the proof will be used in the algorithm for solving (15).

**Theorem III.3** *Suppose that $Y^\star$ is the optimal solution for SDP problem* (35), *and one of the following two inequality conditions*

$$\text{tr}(A_1 Y^\star) = \text{tr}(\tilde{C}X^\star) < \Delta_0 \quad (43)$$

*and*

$$\text{tr}(A_3 Y^\star) = \text{tr}(QQ^H X^\star) - 2\Re(a_0^H QQ^H x^\star) + a_0^H QQ^H a_0 < \epsilon \quad (44)$$

*is satisfied. Then SDP problem* (35) *has rank-one solution that can be found in polynomial time.*

*Proof:* Let $b_2 = \text{tr}(A_2 Y^\star)$, which lies in the interval $[N(1-\eta_1), N(1+\eta_2)]$. Hence, SDP problem (35) is equivalent to the following SDP problem:

$$\begin{aligned}
\underset{Y}{\text{minimize}} \quad & \text{tr}(A_0 Y) \\
\text{subject to} \quad & \text{tr}(A_1 Y) \leq \Delta_0 \\
& \text{tr}(A_2 Y) = b_2 \\
& \text{tr}(A_3 Y) \leq \epsilon \\
& \text{tr}(A_4 Y) = 1 \\
& Y \succeq 0.
\end{aligned} \quad (45)$$

Further, SDP problem (45) can be recast equivalently into:

$$\begin{aligned}
\underset{Y}{\text{minimize}} \quad & \text{tr}(A_0 Y) \\
\text{subject to} \quad & \text{tr}(A_1' Y) \leq 0 \\
& \text{tr}(A_2' Y) = 0 \\
& \text{tr}(A_3' Y) \leq 0 \\
& \text{tr}(A_4 Y) = 1 \\
& Y \succeq 0,
\end{aligned} \quad (46)$$

where
$$A'_1 = \begin{bmatrix} \tilde{C} & 0 \\ 0 & -\Delta_0 \end{bmatrix}, A'_2 = \begin{bmatrix} I & 0 \\ 0 & -b_2 \end{bmatrix}, \quad (47)$$
and
$$A'_3 = \begin{bmatrix} QQ^H & -QQ^H a_0 \\ -a_0^H QQ^H & a_0^H QQ^H a_0 - \epsilon \end{bmatrix}. \quad (48)$$
Note that
$$v^\star((35)) = v^\star((45)) = v^\star((46)), \quad (49)$$
and that $Y^\star$ is optimal for (45) and (46). Condition (43) implies the following inequality $\text{tr}(A'_1 Y^\star) < 0$, while inequality (44) means that $\text{tr}(A'_3 Y^\star) < 0$. In other words, one of the inequality constraints in (46) evaluated at $Y^\star$ is strict. Therefore, it follows from [29, Theorem 6.5] that the following QCQP problem:

$$\begin{aligned}
\underset{y}{\text{minimize}} \quad & \text{tr}(A_0 y y^H) \\
\text{subject to} \quad & \text{tr}(A_1 y y^H) \leq \Delta_0 \\
& \text{tr}(A_2 y y^H) = b_2 \\
& \text{tr}(A_3 y y^H) \leq \epsilon \\
& \text{tr}(A_4 y y^H) = 1
\end{aligned} \quad (50)$$

is solvable in polynomial time through solving the relaxed SDP problem (46). Accordingly, the optimal values satisfy
$$v^\star((45)) = v^\star((46)) = v^\star((50)), \quad (51)$$
and (46) or (45) has a rank-one solution, say $y^\star y^{\star H}$ (where $y^\star$ is optimal for (50)). It follows from (49) and (51) that
$$v^\star((35)) = v^\star((45)) = v^\star((46)) = v^\star((50)). \quad (52)$$

Since
$$\text{tr}(A_0 Y^\star) = v^\star((35)) = v^\star((45)) = v^\star((46)) = \text{tr}(A_0 y^\star y^{\star H}), \quad (53)$$
the rank-one solution $y^\star y^{\star H}$ is optimal for the SDP problem (35). This implies that the SDP relaxation is tight, that is,
$$v^\star((30)) = v^\star((35)), \quad (54)$$
and $y^\star$ is the optimal solution for problem (30).

Note additionally that (34), (52) and (54) yield
$$v^\star((15)) = v^\star((30)) = v^\star((35)) = v^\star((45)) = v^\star((46)) = v^\star((50)). \quad (55)$$
Also, problems (35), (45) and (46) share the optimal solutions $Y^\star$ and $y^\star y^{\star H}$, and problems (30) and (50) share the optimal solution $y^\star$.

Now we present how to construct the rank-one optimal solution $y^\star y^{\star H}$ for relaxed SDP problem (46) (or equivalently problem (45)).

Suppose that condition (43) is true (i.e. $\text{tr}(A'_1 Y^\star) < 0$). Employing the rank-one matrix decomposition lemma (Lemma (III.1)), we obtain $Y^\star = \sum_{k=1}^R y_k y_k^H$ ($R$ is the rank of $Y^\star$) such that
$$\text{tr}(A'_2 y_k y_k^H) = 0, \quad \text{tr}(A'_3 y_k y_k^H) \leq 0, k = 1, \ldots, R, \quad (56)$$
i.e., we perform $\mathcal{D}_1(Y^\star, A'_2, A'_3)$.

Due to conditions (56), we can conclude that there exists at least one vector $y_l$, $l \in \{1, \ldots, R\}$, such that
$$\text{tr}(A'_1 y_l y_l^H) < 0, \quad (57)$$
and $y_l = [x_l; t_l]$ fulfills $t_l \neq 0$ (cf. the third paragraph in the proof of [29, Theorem 6.5]).

Let $\bar{y} = [x_l/t_l; 1]$. It can be easily verified that $\bar{y}\bar{y}^H$ is not only feasible, but also optimal for problem (46) (cf. complementary condition (37)). Then it follows from (53) that $\bar{y}\bar{y}^H$ is optimal for SDP problem (35), and from (55) that $\bar{y}$ is optimal for problem (30) and $x_l/t_l$ is optimal for original problem (15).

Suppose that condition (44) holds (i.e. $\text{tr}(A'_3 Y^\star) < 0$). In a similar way, we can conduct the decomposition $\mathcal{D}_1(Y^\star, A'_1, A'_2)$, returning $Y^\star = \sum_{k=1}^R y_k y_k^H$ such that
$$\text{tr}(A'_1 y_k y_k^H) \leq 0, \text{tr}(A'_2 y_k y_k^H) = 0, k = 1, \ldots, R. \quad (58)$$

Select $y_l = [x_l; t_l]$ with $t_l \neq 0$ from $\{y_1 y_1^H, \ldots, y_R y_R^H\}$ such that $\text{tr}(A'_3 y_l y_l^H) < 0$. Then $\bar{y}\bar{y}^H$ is optimal for SDP problem (35) with $\bar{y} = [x_l/t_l; 1]$. Therefore, the vector $\bar{y} = [x_l/t_l; 1]$ is optimal for problem (30) and $x_l/t_l$ is optimal for original problem (15). The proof is thus complete. ■

Using Theorem III.3, Algorithm 2 summarizes the procedure (as stated in the proof) for finding the globally optimal solution of problem (15), under condition (43) or (44) or both.

---

**Algorithm 2** Procedure for Solving QCQP Problem (15)

**Input:** $\hat{R}, \tilde{C}, Q, a_0, \Delta_0, \eta_1, \eta_2, \epsilon$;
**Output:** An optimal solution $a^\star$ of problem (15);
1: define $A_i$ as in (31)-(33), solve SDP (35) finding $Y^\star$ as in (42), and define $A'_i$ as in (47)-(48);
2: if $\text{tr}(A_1 Y^\star) < \Delta_0$, then implement the matrix decomposition $\{y_i\} = \mathcal{D}_1(Y^\star, A'_2, A'_3)$ and pick up $y_l = [x_l; t_l]$ with nonzero $t_l$ such that $\text{tr}(A'_1 y_l y_l^H) < 0$; go to step 4;
3: if $\text{tr}(A_3 Y^\star) < \epsilon$, then perform the decomposition $\{y_i\} = \mathcal{D}_1(Y^\star, A'_1, A'_2)$ and select $y_l = [x_l; t_l]$ with nonzero $t_l$ such that $\text{tr}(A'_3 y_l y_l^H) < 0$;
4: output $a^\star = x_l/t_l$.

---

The computational cost in Algorithm 2 is dominated by the cost of solving SDP problem (35).

Remark that the proof of Theorem III.3 indeed shows how [29, Theorem 6.5] is applied to SDP problem (35) with the double-sided constraint. It also implies the key that the double-sided constraint is accounted as an equality constraint (as in (45)); otherwise, if the double-sided constraint is treated as two inequality constraints, then [29, Theorem 6.5] is not applicable.

In the following analysis, we only focus on the scenario that $Y^\star$ (as defined in (42)) is such that:
$$\text{tr}(A_3 Y^\star) = \text{tr}(QQ^H X^\star) - 2\Re(a_0^H QQ^H x^\star) + a_0^H QQ^H a_0 = \epsilon. \quad (59)$$

Under condition (59), it is still possible to find a mild sufficient condition under which SDP problem (35) has a rank-one solution. The following theorem establishes such conditions.

**Theorem III.4** *Suppose that $Y^\star$ (as defined in (42)) is optimal for SDP problem (35). Suppose that condition (59) holds true. If the number $a_0^H QQ^H x^\star$, which is in general a complex number, is neither a positive number nor zero (i.e., $a_0^H QQ^H x^\star \not\geq 0$), then SDP problem (35) has a rank-one solution.*

*Proof:* Let $a_0^H QQ^H x^\star = re^{j\theta}$. Then it follows that $r \neq 0$, $\theta \in (0, 2\pi)$ and
$$\Re(a_0^H QQ^H x^\star) < r \quad (60)$$
due to the assumption about $a_0^H QQ^H x^\star$.

Define a new matrix
$$\bar{Y} = \begin{bmatrix} X^\star & e^{-j\theta} x^\star \\ e^{j\theta} x^{\star H} & 1 \end{bmatrix}. \quad (61)$$

It can be immediately seen that $\bar{Y}$ is positive semidefinte and
$$\text{tr}(A_i Y^\star) = \text{tr}(A_i \bar{Y}), i = 0, 1, 2, 4. \quad (62)$$

However, considering (59) and (60), we can also see that

$$\text{tr}(A_3 \bar{Y}) = \text{tr}(QQ^H X^\star) - 2r + a_0^H QQ^H a_0 < \epsilon. \quad (63)$$

This implies that $\bar{Y}$ is a new optimal solution for (35), with $\text{tr}(A_3 \bar{Y}) < \epsilon$. It then follows from Theorem III.3 that SDP (35) has a rank-one solution. The proof is complete. ∎

From the proof of Theorem III.4, it can be seen that if $a_0^H QQ^H x^\star \not\geq 0$, then we still can apply Algorithm 2 with $Y^\star = \bar{Y}$ as defined in (61), in order to find the optimal solution for original problem (15).

The following stronger rank-one matrix decomposition lemma can also be used for the purpose of obtaining a rank-one solution for the considered SDP problems.

**Lemma III.5 (Theorem 2.3 in [30])** *Let $X$ be a non-zero $N \times N$ ($N \geq 3$) complex Hermitian positive semidefinite matrix of rank $R$, and $A_i$, $i = 1, 2, 3, 4$ be Hermitian matrices. Suppose that $(\text{tr}(A_1 Y), \text{tr}(A_2 Y), \text{tr}(A_3 Y), \text{tr}(A_4 Y)) \neq (0, 0, 0, 0)$ for any non-zero complex Hermitian positive semidefinite matrix $Y$. Then the following two statements are true.*

1) *If $R \geq 3$, then one can find with polynomial time complexity a vector $x \in \text{Range}(X)$ such that*

$$x^H A_i x = \text{tr}(A_i X), \quad i = 1, 2, 3, 4.$$

*(synthetically denoted as $x = \mathcal{D}_2(X, A_1, A_2, A_3, A_4)$).*

2) *If $R = 2$, then for any $z \notin \text{Range}(X)$, there exists $x \in \mathbb{C}^N$ in the linear subspace spanned by $z$ and $\text{Range}(X)$, such that $x^H A_i x = \text{tr}(A_i X)$, $i = 1, 2, 3, 4$.*

Therefore, one more sufficient condition ensuring that the solution of SDP problem (35) is of rank one can be derived based on Lemma III.5. This condition is given by the following theorem.

**Theorem III.6** *Suppose that $Y^\star$ (as defined in (42)) is optimal for SDP problem (35). If the rank of $Y^\star$ is three or above, then there is a rank-one solution for SDP problem (35).*

*Proof:* See Appendix A ∎

Remark that if the rank of $Y^\star$ happens to be one, then in fact $y^\star$ with $Y^\star = y^\star y^{\star H}$ is the solution for problem (30). However, if $Y^\star$ is of rank three or above, one can apply the procedure described in the proof of Theorem III.6 to solve problem (30).

We remark that the case where $Y^\star$ is of rank two rarely occurs in extensive numerical simulations. However if it occurs, SDP problem (35) may not have a rank-one solution. Nonetheless, it is possible to apply the second statement of Lemma III.5 in order to find an approximate solution for SDP problem (35). In other words, we can randomly pick up a vector $y'$ outside $\text{Range}(Y^\star)$ (so that the dimension of $\mathcal{S} = \text{Span}(\text{Range}(Y^\star) \cup \{y'\})$ is three), and perform $\mathcal{D}_2(Y^\star, A_1, A_2, A_3, A_4)$ in the span $\mathcal{S}$, in order to obtain such vector $\tilde{y} \in \mathcal{S}$ that $\tilde{y}^H A_1 \tilde{y} = b_1$, $\tilde{y}^H A_2 \tilde{y} = \Delta_1$, $\tilde{y}^H A_3 \tilde{y} = \epsilon$, and $\tilde{y}^H A_4 \tilde{y} = 1$. This implies that $\tilde{y} \tilde{y}^H$ is feasible for problem (35), but the global optimality is not guaranteed. Therefore, $\tilde{y}$ is a suboptimal/approximate solution for problem (30).

Finally, let us summarize the solvable cases for problem (30), i.e., the cases when the globally optimal solution for (35) is guaranteed to be rank-one. Suppose $Y^\star$ (as defined in (42)) is a general-rank solution for SDP problem (35). If one of the following conditions is satisfied, then a rank-one optimal solution for SDP problem (35) can be found in polynomial time:

1) Rank $(Y^\star) = 1$ or Rank $(Y^\star) \geq 3$;
2) Rank $(Y^\star) = (\geq)2$ and $\text{tr}(A_1 Y^\star) < \Delta_0$;
3) Rank $(Y^\star) = (\geq)2$ and $\text{tr}(A_3 Y^\star) < \epsilon$;
4) Rank $(Y^\star) = (\geq)2$, $\text{tr}(A_3 Y^\star) = \epsilon$, and $a_0^H QQ^H x^\star \not\geq 0$.

## IV. A NEW QUADRATIC CONSTRAINT FOR THE DESIRED STEERING VECTOR

Aiming to further improve the performance of the robust adaptive beamformer, we herein propose a new quadratic constraint, which forces the signal of interest steering vector to separate itself from those vectors with DOAs in the complement of $\Theta$, including the linear combinations of all interference steering vectors.

Rather than exploiting matrix $\tilde{C}$ in (9), we analyze the following matrix

$$C = \int_\Theta d(\theta) d^H(\theta) d\theta. \quad (64)$$

Here the difference lies in the integral interval, which is $\Theta$ for the matrix $C$ and the complement of $\Theta$ for the matrix $\tilde{C}$. Then, let us consider the following constraint

$$a^H C a \geq \Delta_1, \quad (65)$$

where

$$\Delta_1 = \min_{\theta \in \Theta} d^H(\theta) C d(\theta). \quad (66)$$

Through the benchmark line $\Delta_1$, it is expected that if $d^H(\theta) C d(\theta) \geq \Delta_1$ for $\theta \in [-90°, 90°]$, then $\theta \in \Theta$ is a possible direction of the signal of interest steering vector. Otherwise, $\theta \in \tilde{\Theta}$ is a direction of steering vector of no interest (or a possible interference direction). In order to illustrate the effectiveness of $\Delta_1$ and $\Delta_0$ (computed by (10)), we draw in Fig. 1 two figures for the desired sector $\Theta = [0°, 10°]$: one for $d^H(\theta) C d(\theta)$ with $\Delta_1$ and the other for $d^H(\theta) \tilde{C} d(\theta)$ with $\Delta_0$ (here subfigure Fig. 1.(b) is the same as in [18, Fig. 2]).

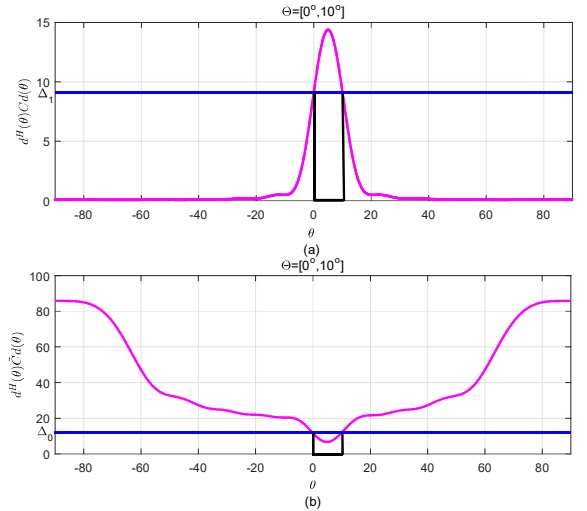

Fig. 1. Two benchmark lines $\Delta_0$ and $\Delta_1$ with the angular sector $\Theta = [0°, 10°]$; (a) for $\theta \in \Theta$, $d^H(\theta) C d(\theta) \geq \Delta_1$ (b) for $\theta \in \Theta$, $d^H(\theta) \tilde{C} d(\theta) \leq \Delta_0$

Capitalizing on the quadratic constraint, we can formulate several robust adaptive beamforming problems based on signal of interest steering vector estimation, which can be viewed as alternative designs to the designs introduced above, aiming to improve the output SINR and/or the output power. First, consider the following steering vector

estimation problem:

$$\begin{aligned}\underset{\boldsymbol{a}}{\text{minimize}} \quad & \boldsymbol{a}^H \hat{\boldsymbol{R}}^{-1} \boldsymbol{a} \\ \text{subject to} \quad & \boldsymbol{a}^H \boldsymbol{C} \boldsymbol{a} \geq \Delta_1 \\ & \|\boldsymbol{a}\|^2 = N,\end{aligned} \qquad (67)$$

where the prior knowledge required is as little as in (8); namely the desired angular sector and the antenna array geometry. It is a two-constraint homogenous QCQP problem, which is known to be solvable, by the same method developed for solving (8) (again, for example see [29]). Thus, it is possible to modify the approach in [18] to solve problem (67).

We also consider the following steering vector estimation problem:

$$\begin{aligned}\underset{\boldsymbol{a}}{\text{minimize}} \quad & \boldsymbol{a}^H \hat{\boldsymbol{R}}^{-1} \boldsymbol{a} \\ \text{subject to} \quad & \boldsymbol{a}^H \boldsymbol{C} \boldsymbol{a} \geq \Delta_1 \\ & N(1-\eta_1) \leq \|\boldsymbol{a}\|^2 \leq N(1+\eta_2) \\ & \|\boldsymbol{Q}^H(\boldsymbol{a} - \boldsymbol{a}_0)\|^2 \leq \epsilon.\end{aligned} \qquad (68)$$

When $\boldsymbol{Q} = \boldsymbol{I}$, new beamforming problem is the following QCQP (similar completely to (16)):

$$\begin{aligned}\underset{\boldsymbol{a}}{\text{minimize}} \quad & \boldsymbol{a}^H \hat{\boldsymbol{R}}^{-1} \boldsymbol{a} \\ \text{subject to} \quad & \boldsymbol{a}^H \boldsymbol{C} \boldsymbol{a} \geq \Delta_1 \\ & N(1-\eta_1) \leq \|\boldsymbol{a}\|^2 \leq N(1+\eta_2) \\ & \|\boldsymbol{a} - \boldsymbol{a}_0\|^2 \leq \epsilon.\end{aligned} \qquad (69)$$

Observe the first constraint in (68) is equivalent to

$$\boldsymbol{a}^H(-\boldsymbol{C})\boldsymbol{a} \leq -\Delta_1, \qquad (70)$$

and Algorithms 1 and 2 can be applied to solve (68) for $\boldsymbol{Q} = \boldsymbol{I}$ and for general $\boldsymbol{Q}$, respectively, with $\tilde{\boldsymbol{C}} := -\boldsymbol{C}$ and $\Delta_0 := -\Delta_1$. However, often $\tilde{\boldsymbol{C}}$ is PSD and $\Delta_0$ is a positive number. For the ease of understanding the physical meaning, we present the new algorithms to (68) in a short way while keeping the form of $\boldsymbol{a}^H \boldsymbol{C} \boldsymbol{a} \geq \Delta_1$.

Let us first solve (69), namely, $\boldsymbol{Q} = \boldsymbol{I}$ in (68). We have the following theorem that states the guaranty of the global optimality for problem (69):

**Theorem IV.1** *QCQP problem* (69) *is solvable and an optimal solution can be constructed tractably within polynomial time complexity.*

The proof is similar to that of Theorem III.2. The procedure to output an optimal solution also includes solving its SDP relaxation problem

$$\begin{aligned}\underset{\boldsymbol{X},\boldsymbol{y}}{\text{minimize}} \quad & \text{tr}(\hat{\boldsymbol{R}}^{-1}\boldsymbol{X}) \\ \text{subject to} \quad & \text{tr}(\boldsymbol{C}\boldsymbol{X}) \geq \Delta_1 \\ & N(1-\eta) \leq \text{tr}\,\boldsymbol{X} \leq N(1+\eta) \\ & \text{tr}\left(\begin{bmatrix} \boldsymbol{I} & -\boldsymbol{a}_0 \\ -\boldsymbol{a}_0^H & N \end{bmatrix} \begin{bmatrix} \boldsymbol{X} & \boldsymbol{y} \\ \boldsymbol{y}^H & 1 \end{bmatrix}\right) \leq \epsilon, \\ & \begin{bmatrix} \boldsymbol{X} & \boldsymbol{y} \\ \boldsymbol{y}^H & 1 \end{bmatrix} \succeq \boldsymbol{0},\end{aligned} \qquad (71)$$

followed by some postprocessing steps. The algorithm for finding the optimal solution for (69) is summarized into Algorithm 3. The computational complexity of the algorithm is dominated by solving SDP problem (71).

Now we solve problem (68) with general $\boldsymbol{Q}$. For beamforming problem (68), the SDP relaxation problem is similar to (35), and it

**Algorithm 3** Procedure for Solving QCQP Problem (69)

`Input:` $\hat{\boldsymbol{R}}$, $\boldsymbol{C}$, $\boldsymbol{a}_0$, $\Delta_1$, $\eta$, $\epsilon$;
`Output:` An optimal solution $\boldsymbol{a}^\star$ of problem (69);
1: solve its SDP relaxation problem (71) finding $(\boldsymbol{X}^\star, \boldsymbol{y}^\star)$;
2: perform the rank-one matrix decomposition $\boldsymbol{x} = \mathcal{D}_1(\boldsymbol{X}^\star, \boldsymbol{C} - \frac{b_1}{b_2}\boldsymbol{I}, \boldsymbol{a}_0\boldsymbol{a}_0^H - \frac{b_3}{b_2}\boldsymbol{I})$, where $b_1 = \text{tr}(\boldsymbol{C}\boldsymbol{X}^\star)$, $b_2 = \text{tr}\,\boldsymbol{X}^\star$ and $b_3 = \text{tr}(\boldsymbol{a}_0\boldsymbol{a}_0^H\boldsymbol{X}^\star)$;
3: set $\boldsymbol{x}^\star = \sqrt{b_2}\frac{\boldsymbol{x}}{\|\boldsymbol{x}\|}$;
4: output $\boldsymbol{a}^\star = \boldsymbol{x}^\star e^{-j\arg(\boldsymbol{a}_0^H \boldsymbol{x}^\star)}$.

is given as:

$$\begin{aligned}\underset{\boldsymbol{Y}}{\text{minimize}} \quad & \text{tr}(\boldsymbol{A}_0 \boldsymbol{Y}) \\ \text{subject to} \quad & \text{tr}(\bar{\boldsymbol{A}}_1 \boldsymbol{Y}) \geq \Delta_1 \\ & N(1-\eta_1) \leq \text{tr}(\boldsymbol{A}_2 \boldsymbol{Y}) \leq N(1+\eta_2) \\ & \text{tr}(\boldsymbol{A}_3 \boldsymbol{Y}) \leq \epsilon \\ & \text{tr}(\boldsymbol{A}_4 \boldsymbol{Y}) = 1 \\ & \boldsymbol{Y} \succeq \boldsymbol{0},\end{aligned} \qquad (72)$$

where

$$\bar{\boldsymbol{A}}_1 = \begin{bmatrix} \boldsymbol{C} & \boldsymbol{0} \\ \boldsymbol{0} & 0 \end{bmatrix}. \qquad (73)$$

To find a globally optimal solution for (68), the following theorem similar to Theorem III.3 can be established.

**Theorem IV.2** *Suppose that $\boldsymbol{Y}^\star$ is (as defined in* (42)*) is an optimal solution for* (72)*, and one of the following inequalities*

$$\text{tr}(\bar{\boldsymbol{A}}_1 \boldsymbol{Y}^\star) = \text{tr}(\boldsymbol{C}\boldsymbol{X}^\star) > \Delta_1 \qquad (74)$$

*and* (44) *is satisfied. Then SDP problem* (72) *has a rank-one solution and the solution can be constructed within polynomial time complexity.*

Furthermore, in order to guarantee that SDP problem (72) has a rank-one solution, the results similar to Theorems III.4 and III.6 can be shown to remain true, since they are sensitive only to the number of constraints. Similar to Algorithm 2, we summarize the procedure for solving (68) into Algorithm 4.

**Algorithm 4** Procedure for Solving QCQP Problem (68)

`Input:` $\hat{\boldsymbol{R}}$, $\boldsymbol{C}$, $\boldsymbol{Q}$, $\boldsymbol{a}_0$, $\Delta_1$, $\eta_1$, $\eta_2$, $\epsilon$;
`Output:` An optimal solution $\boldsymbol{a}^\star$ of problem (68);
1: define $\bar{\boldsymbol{A}}_1' = [\boldsymbol{C}, \boldsymbol{0}; \boldsymbol{0}, -\Delta_1]$, solve SDP (72) finding $\boldsymbol{Y}^\star$ as in (42), and define $\boldsymbol{A}_2'$ and $\boldsymbol{A}_3'$ as in (47) and (48), respectively;
2: if $\text{tr}(\bar{\boldsymbol{A}}_1 \boldsymbol{Y}^\star) > \Delta_1$, then implement the matrix decomposition $\{\boldsymbol{y}_i\} = \mathcal{D}_1(\boldsymbol{Y}^\star, \boldsymbol{A}_2', \boldsymbol{A}_3')$ and pick up $\boldsymbol{y}_l = [\boldsymbol{x}_l; t_l]$ with nonzero $t_l$ such that $\text{tr}(\bar{\boldsymbol{A}}_1' \boldsymbol{y}_l \boldsymbol{y}_l^H) > 0$; go to step 4;
3: if $\text{tr}(\boldsymbol{A}_3 \boldsymbol{Y}^\star) < \epsilon$, then perform the decomposition $\{\boldsymbol{y}_i\} = \mathcal{D}_1(\boldsymbol{Y}^\star, \bar{\boldsymbol{A}}_1', \boldsymbol{A}_2')$ and select $\boldsymbol{y}_l = [\boldsymbol{x}_l; t_l]$ with nonzero $t_l$ such that $\text{tr}(\boldsymbol{A}_3' \boldsymbol{y}_l \boldsymbol{y}_l^H) < 0$;
4: output $\boldsymbol{a}^\star = \boldsymbol{x}_l/t_l$.

## V. NUMERICAL EXAMPLES

In this section, we present several numerical examples aiming to evaluate the performance of the proposed MVDR robust adaptive beamformers.

## A. Example 1: Signal Look Direction Mismatch

Consider a uniform linear array with $N = 12$ omni-directional antenna elements spaced half a wavelength apart of each other. The array noise is a spatially and temporally white Gaussian vector with zero mean and covariance $\boldsymbol{I}$. Two interferers with the same interference-to-noise ratio (INR) of 30 dB are assumed to impinge upon the array from the angles $\theta_1 = -15°$ and $\theta_2 = 15°$ with respect to the array broadside, and the desired signal is always present in the training data cell. The training sample size $T$ is preset to 100. The angular sector $\Theta$ of interest is $[0°, 10°]$, and the presumed direction is assumed to be $\theta_0 = 5°$ ($\theta_0$ is the middle point of $\Theta$, and thus $\boldsymbol{a}_0 = \boldsymbol{d}(\theta_0)$ is given by default). The actual signal impinges upon the array from direction $\theta = 7°$.

The norm perturbation parameters $\eta_1$ and $\eta_2$ for the proposed beamformers are both set to 0.5, and the similarity constraint parameter $\epsilon = 0.3N$ in both (16) and (69). All results are averaged over 200 simulation runs. In each run, three problems (8), (16), and (69) are solved for different SNR= $[-10, 60]$ dB. The beamformers resulted from problems (8), (16), and (69) are respectively termed "KVH Beamformer", "New Beamformer 1", and "New Beamformer 2". We evaluate the beamformer performance in terms of the array output SINR as well as array output power.

Fig. 2 demonstrates the beamformer output SINR versus the SNR. As we can see, the beamformer output SINR obtained through (69) is higher than that through (16), especially at moderate and high SNR. This means that the new quadratic constraint (65) leads to some performance gain, comparing with the quadratic constraint (11). Observe that the both new beamformers have higher SINR than the KVH Beamformer has, especially moderate SNR. This implies that additionally considering the similarity constraint in the beamformer designs does yield some gain.

In Fig. 3, assuming an SNR of 30dB, we plot the array output SINR versus the number of snapshots. We observe that the performance gap between the KVH beamformer and any one of the two new beamformers is clear, which means the new designs do have some gains in terms of the output SINR. It is also seen that the output SINR by New Beamformer 2 is higher than that by New Beamformer 1, especially when $T$ increases.

Fig. 4 shows the output power versus SNR for the beamformers tested. As can be seen, the output power of the three beamformers increases as the SNR increases. Also, the performance of the two new beamformers is better than that of the KVH beamformer, especially in high SNR region.

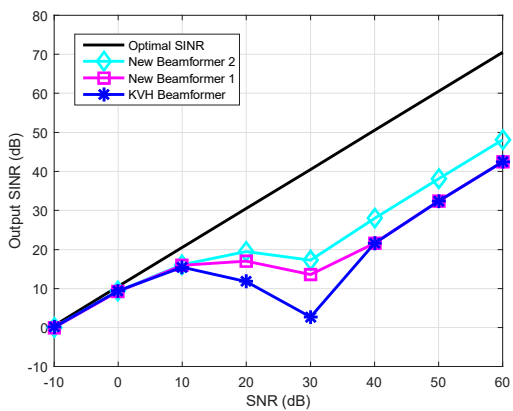

Fig. 2. Average beamformer output SINR versus SNR, with $T = 100$.

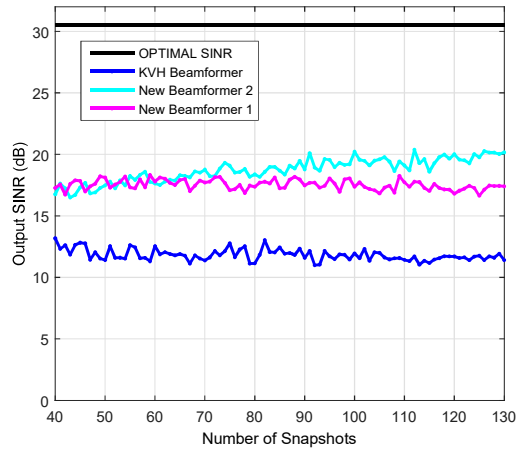

Fig. 3. Average beamformer output SINR versus training sample size $T$, with SNR=30dB.

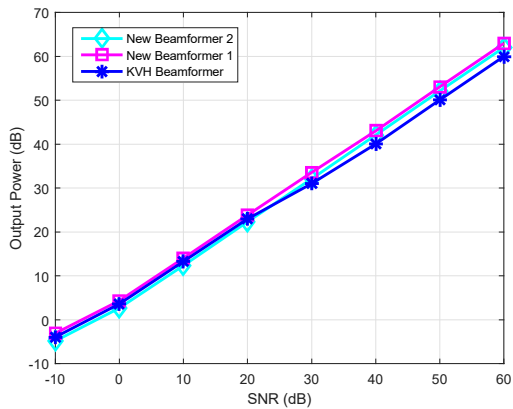

Fig. 4. Average array output power versus SNR, with $T = 100$

## B. Example 2: Waveform Distortion in Inhomogeneous Medium

In this example, we take into account mismatch caused also by wavefront distortion in an inhomogeneous medium [18], with no signal look direction mismatch (i.e. setting the nominal direction $\theta_0 = 9°$ and the actual direction $\theta = 9°$). Specifically, we assume that the signal steering vector is distorted by wave propagation effects in the way that independent-increment phase distortions are accumulated by the components of the steering vector, and assume that the phase increments are independent Gaussian variables each with zero mean and standard deviation 0.01, and they are randomly generated and remain unaltered in each simulation run. All other settings are the same as those in Example 1.

Fig. 5 shows the array output SINR versus the SNR for the three beamformers tested in the previous numerical example. We again can observe that the output SINR by (69) (New Beamformer 2) is bigger than that by (16) (New Beamformer 1), which clearly is larger than the beamformer based on (8) (the KVH Beamformer), especially at moderate SNR.

## C. Example 3: Beamforming Based on An Ellipsoid Constraint (Beyond the Similarity Constraint)

In this example, we study the performance of Algorithms 2 and 4 proposed for the beamforming problems with an ellipsoidal constraint (beyond the similarity constraint). We adopt the method in [7] to

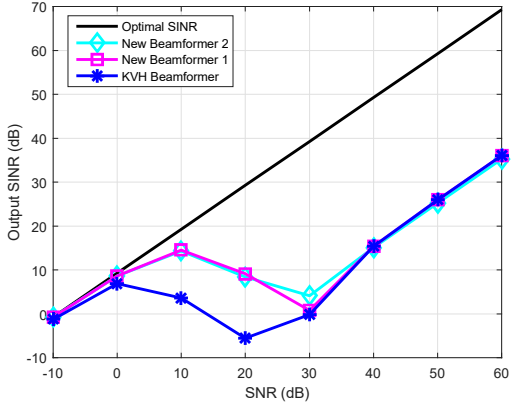

Fig. 5. Average beamformer output SINR versus SNR, with $T = 100$.

generate an ellipsoid

$$\mathcal{E} = \{\boldsymbol{a} \mid (\boldsymbol{a} - \boldsymbol{a}_0)^H \boldsymbol{P}^{-1} (\boldsymbol{a} - \boldsymbol{a}_0) \leq \epsilon\}, \quad (75)$$

where the actual steering vector is located. In other words, $\boldsymbol{Q}\boldsymbol{Q}^H = \boldsymbol{P}^{-1}$. We collect $L = 100$ equally spaced samples at the angle sector $\Theta = [0°, 10°]$, to which the direction of the signal of interest belongs. Then the center $\boldsymbol{a}_0$ and the matrix $\boldsymbol{P}$ are, respectively, the sample mean and the sample covariance matrix of different steering vectors with angles in the sector. In other words,

$$\boldsymbol{a}_0 := \bar{\boldsymbol{a}} = \frac{1}{L} \sum_{l=1}^{L} \boldsymbol{a}(\theta_l)$$

and

$$\boldsymbol{P} = \frac{1}{L} \sum_{l=1}^{L} (\boldsymbol{a}(\theta_l) - \bar{\boldsymbol{a}})(\boldsymbol{a}(\theta_l) - \bar{\boldsymbol{a}})^H,$$

where

$$\theta_l = \frac{\theta_{\min} + \theta_{\max}}{2} + \left(-\frac{1}{2} + \frac{l-1}{L-1}\right)(\theta_{\max} - \theta_{\min}), \, l = 1, \ldots, L.$$

In order to guarantee that $\boldsymbol{P}$ is positive definite in our cases, let

$$\boldsymbol{P} := \boldsymbol{P} + 0.1\boldsymbol{I}.$$

The parameter $\epsilon$ of the ellipsoid takes value of $0.45N$ in both (15) and (68).

We test the performance of beamformers associated with (15) and (68) both including the ellipsoidal constraint, as well as with (16) and (69) both containing the similarity constraint. In this example, for the all proposed beamformers, $\boldsymbol{a}_0$ is the sample mean, rather than $\boldsymbol{a}_0 = \boldsymbol{d}(\theta_0)$ (as in Examples 1 and 2). All other parameters are the same as those in Example 2. We call "New Beamformer 3" and "New Beamformer 4" for the beamformers designed by (15) and (68) both with an ellipsoidal constraint, respectively, and again "New Beamformer 1" and "New Beamformer 2" for (16) and (69) both with a similarity constraint, respectively.

Fig. 6 depicts the average output SINR versus the SNR. It can be seen from the figure that the performance of New Beamformer 4 is slightly better than that of New Beamformer 3. Observe that the performance of New Beamformer 2 is better than that of New Beamformer 1, which in turn is better than the KVH Beamformer. The two observations mean again that the new quadratic constraint (65) is better than the quadratic constraint (11) in terms of the SINR performance. We further observe that at SNR=30 dB, New Beamformer 2 has the best performance over the four new beamformers, however, in the higher SNR region, the output SINR by New Beamformer 2 is slightly worse than those by New Beamformers 4 and 3. This means that the performance of New Beamformer 2 is more or less equal to that of New Beamformer 4, and New Beamformer 3 is clearly better than the performance of New Beamformer 1, which implies that the replacement of the similarity constraint by the ellipsoidal constraint leads to some SINR gain.

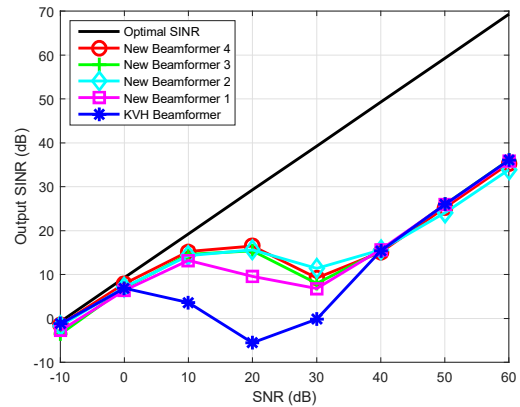

Fig. 6. Average beamformer output SINR versus SNRs, with $T = 100$.

### D. Example 4: The Sector of Interest Directional Dependence (Sector of Interest is Far Away from the Broadside)

In this example, the angular sector of interest is $\Theta = [50°, 60°]$, i.e., it is far enough from the broadside of the antenna array. Thus, the presumed direction is $\theta_0 = 55°$ (the middle point of $\Theta$), while the actual signal direction is also $\theta = 55°$, i.e., no signal look direction mismatch is assumed. Two interferers impinge on the array from angles $\theta_1 = 25°$ and $\theta_2 = 85°$. The mismatch caused by wavefront distortion is considered, and every phase increment is an independent Gaussian variable with zero mean and standard deviation 0.02. All other setups are the same as in Example 2. Moreover, in order to generate the ellipsoidal constraint (see (75)), we set $L = 64$. All other parameters set for the ellipsoid are the same as the corresponding ones in Example 3. The KVH and all four proposed new beamformers are tested.

It can be seen from Fig. 7 that the four proposed beamformers show significant performance superiority comparing with the KVH beamformer in the SNR region of $[-10, 30]$ dB. Thus, it can be concluded that the response for the KVH beamformer is severely dependent on the direction of the sector of interest. If the sector of interest is far enough from the antenna array broadside and is rather close to the antenna end-fire, the quadratic constraint in the KVH beamformer is insufficient to prevent the signal of interest cancellation for insufficiently high SNR. Thus, the KVH beamformer is not invariant to the sector of interest directions, and initial presteering of the antenna array has to be done before applying it to guarantee high performance. For the new proposed beamformers, however, the similarity constraint ensures that the optimal estimate of the steering vector is sufficiently close to the average of the steering vectors with their DOAs inside of the angular sector of interest. It prevents the signal of interest cancellation independent on the sector direction. Thus, all proposed beamformers are invariant to the sector of interest direction, which can be concluded by comparing the results of Examples 3 and 4, in both of which the behaviors of the four proposed beamformers are similar.

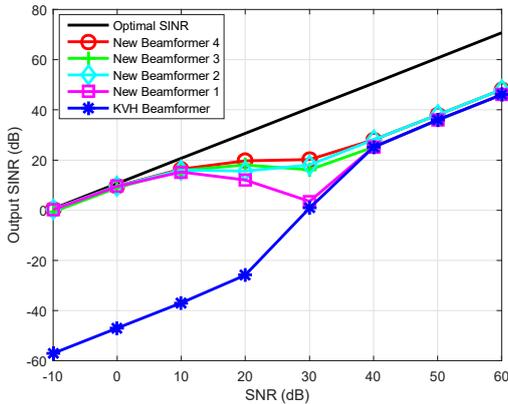

Fig. 7. Average beamformer output SINR versus SNRs, with $\Theta = [50°, 60°]$ and $T = 100$.

## VI. Conclusion

We have considered the MVDR robust adaptive beamforming problem based on optimal steering vector estimation with limited prior knowledge. A new beamformer design problem has been studied, by considering the beamformer output power maximization problem, subject to a relaxed double-sided norm constraint, an additional similarity constraint, and the constraint enforcing the desired signal DOA to be far away from the DOA interval of all linear combinations of the interference steering vectors. It has turned out that the maximization problem is a non-convex QCQP with three constraints, one of which is inhomogeneous and another of which is double-sided. We have shown that the QCQP problem is equivalent to its SDP relaxation problem, and presented how to efficiently get an optimal solution. Furthermore, we have generalized the similarity constraint to an ellipsoidal constraint, and the new formulated problem is hard with no global optimality guaranteed. In that case, we have established sufficient optimality conditions. Besides, a new quadratic constraint on the actual signal steering vectors has been proposed, aiming to ameliorate the array performance. The improved performance of the proposed robust beamformer has been demonstrated by simulations in terms of the output SINR and the output power.

## Appendix A
## Proof of Theorem III.6

*Proof:* Let $b_i = \text{tr}(A_i Y^\star)$, $i = 1, 2, 3, 4$. Therefore, $Y^\star$ is optimal for (45) as well. Capitalizing on the rank-one decomposition Lemma III.5, we perform the decomposition $\mathcal{D}_2(Y^\star, A_1, A_2, A_3, A_4)$, returning a vector $y$ such that $y^H A_i y = b_i$, $i = 1, 2, 3, 4$, and $y \in \text{Range}(Y^\star)$. Notice that $\text{Range}(Y^\star) \subseteq \text{Null}(Z^\star)$ due to the complementary condition (37) for problem (45), where $Z^\star = A_0 - z_1^\star A_1 - (z_0^\star + z_2^\star) A_2 - z_3^\star A_3 - z_4^\star A_4$ for some dual optimal solution. This means that $yy^H$ is not only a feasible solution, but an optimal solution for (45), and then for (35). ∎